\documentclass[twocolumn]{aastex631}

\received{XXX XX, XXXX}
\revised{XXX XX, XXXX}
\accepted{XXX XX, XXXX}
\submitjournal{ApJ}

\shorttitle{PAC. II. Morphology dependence of SHMR}
\shortauthors{Xu \& Jing}

\begin{document}
\defcitealias{2021arXiv210911738X}{Paper I}
\title{Photometric Objects Around Cosmic Webs (PAC) Delineated in a Spectroscopic survey. II. Morphology, Color and Size Dependences of the Stellar-halo Mass Relation for Massive Galaxies}

\correspondingauthor{Yipeng Jing}
\email{ypjing@sjtu.edu.cn}

\author[0000-0002-7697-3306]{Kun Xu}
\affil{Department of Astronomy, School of Physics and Astronomy, Shanghai Jiao Tong University, Shanghai, 200240, People’s Republic of China}

\author[0000-0002-4534-3125]{Yipeng Jing}
\affil{Department of Astronomy, School of Physics and Astronomy, Shanghai Jiao Tong University, Shanghai, 200240, People’s Republic of China}
\affil{Tsung-Dao Lee Institute, and Shanghai Key Laboratory for Particle Physics and Cosmology, Shanghai Jiao Tong University, Shanghai, 200240, People’s Republic of China}



\begin{abstract}
In this paper, we report a robust measurement of the morphology, color and galaxy size dependences of the stellar-halo mass relation (SHMR) at the high mass end ($10^{11.3}{\rm M_{\odot}}<M_{\star}<10^{11.7}{\rm M_{\odot}}$) at redshift $z_s\sim0.6$\footnote{Throughout the paper, we use $z_s$ for redshift, $z$ for the z-band magnitude.}. Applying our method, Photometric objects Around Cosmic webs (PAC),  developed in a previous work to CMASS and HSC-SSP observations, we measure the excess surface density ($\bar{n}_2w_p(r_p)$) of satellites around massive central galaxies with different morphologies indicated by S\'ersic index $n$. We find that more compact (larger $n$) central galaxies are surrounded by more satellites. With the abundance matching method, we estimate halo mass for the central galaxies, and find that halo mass is increased monotonically with $n$, solid evidence for a morphology dependence of SHMR. Specifically, our results show that the most compact galaxies ($n>6$) have the halo mass around 5.5 times larger than the disk galaxies ($n<2$). Similarly, using the effective radius $R_e$ and the rest-frame $u-r$ color, we find that red (large) galaxies reside in halos that are in average $2.6$ ($2.3$) times more massive than those hosting blue (small)  galaxies.
\end{abstract}
\keywords{Galaxy formation(595) --- Galaxy dark matter halos (1880)}

\section{Introduction} \label{sec:intro}
The stellar-halo mass relation (SHMR) is one of the most fundamental relations in galaxy-halo connection (see \citet{2018ARA&A..56..435W} for a review), in which larger dark matter halos host more massive galaxies with a relatively tight scatter. SHMR has been measured with different data and methods in the past decade \citep{2007ApJ...667..760Z,2010MNRAS.404.1111G,2010MNRAS.402.1796W,2012ApJ...752...41Y,2013MNRAS.428.3121M,2019MNRAS.488.3143B}, and it reflects the sophisticated regulation of complex physical processes in galaxy formation such as gas accretion and feedback processes at different mass scales. 

The SHMR is a relation between the halo mass and galactic stellar mass, which is assumed to be independent of other properties of halos or galaxies in early studies. However, recent studies (e.g., \citet{2010MNRAS.402.1942C,2014MNRAS.443.3044Z,2021MNRAS.505.5117Z,2021arXiv210806790Z}) have paid more attention on whether galaxies with different properties have different SHMRs, which is also related to so-called galaxy assembly bias that causes the scatter in the SHMR. Till now, there is no consensus about the dependence on other galaxy properties. For example, some studies found that red central galaxies reside in more massive halos than blue ones \citep{2010MNRAS.402.1942C,2013MNRAS.433..515W,2014MNRAS.443.3044Z, 2015ApJ...799..130R,2015MNRAS.452.1958H,2016MNRAS.457.3200M}, while others found an opposite relation that blue galaxies have more massive host halos \citep{2013ApJ...778...93T, 2018MNRAS.477.1822M, 2019ApJ...871..147G}. Using HI rotation curves, \citet{2019A&A...626A..56P} showed that local massive spiral galaxies are hosted by smaller dark matter halos, implying a morphology dependence of SHMR. Moreover, this dependence does not show up in some of current cosmological hydrodynamical simulations \citep{2020A&A...640A..70M}, while it does exist in some others (e.g., \citet{2021NatAs.tmp..138C}). The galaxy size dependence of SHMR has also been investigated in some studies \citep{2017MNRAS.472.2367C,2017MNRAS.471L..11D,2018MNRAS.473.2714S} but a clear answer to this question does not exist yet. In summary, it is still under hot debate both in observations and in theories if there exist  secondary parameters for the SHMR other than the stellar mass.

In this paper, using the method named Photometric objects Around Cosmic webs (PAC) developed in the first paper of this series \citep{2021arXiv210911738X}, we measure the dependences of SHMR on morphology, galaxy size and color for massive galaxies, and report robust detections of both dependences, which shows unambiguously that large, red or more compact galaxies are located in more massive halos. 

We adopt the cosmology with $\Omega_m = 0.268$, $\Omega_{\Lambda} = 0.732$ and $H_0 = 71{\rm \ km/s/Mpc}$ through out the paper.

\section{Data and method} \label{sec:data}
In \citet[hearafter \citetalias{2021arXiv210911738X}]{2021arXiv210911738X}, we developed a method for estimating the projected density distribution $\bar{n}_2w_p(r_p)$ of photometric objects around spectroscopic objects in a spectroscopic survey, where $\bar{n}_2$ is the mean number density of the photometric objects and $w_p$ is the projected cross-correlation function (PCCF).  This quantity describes the distribution of Photometric sources with certain physical properties (e.g. luminosity, mass, color etc) Around Cosmic webs (PAC) traced by the spectroscopic objects. Since the halo mass is highly correlated with the satellite distribution \citep{2021arXiv210405355W}, PAC is very powerful for studying SHMR.  For details of the method, we refer the readers to \citetalias{2021arXiv210911738X}.

\begin{figure*}
    \plottwo{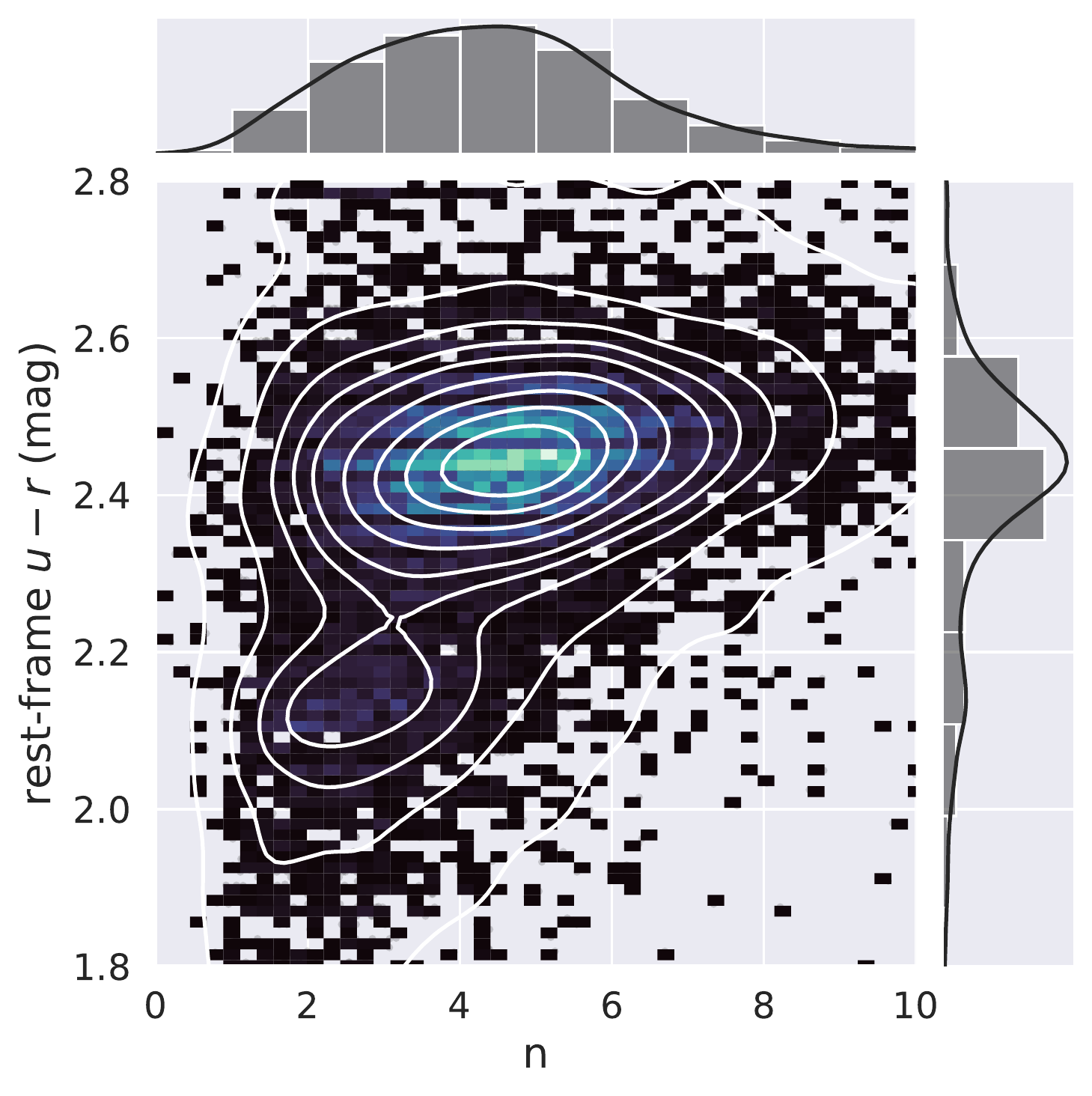}{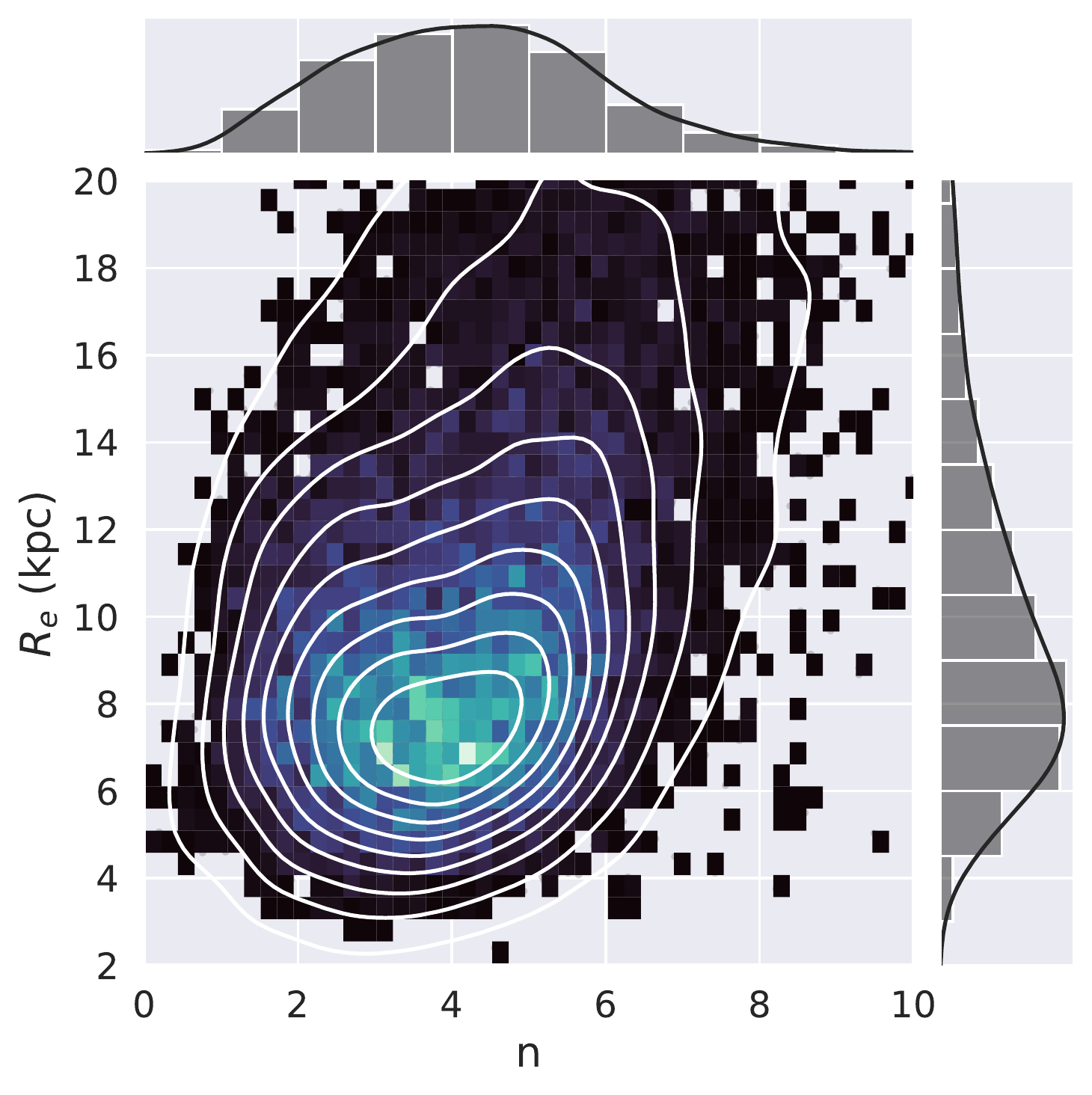}
    \caption{Rest-frame $u-r$ color vs. morphology diagram (left) and $R_e$ vs. morphology diagram (right) of the CMASS sample.}
    \label{fig:f1a}
\end{figure*}

\begin{figure*}
    \plotone{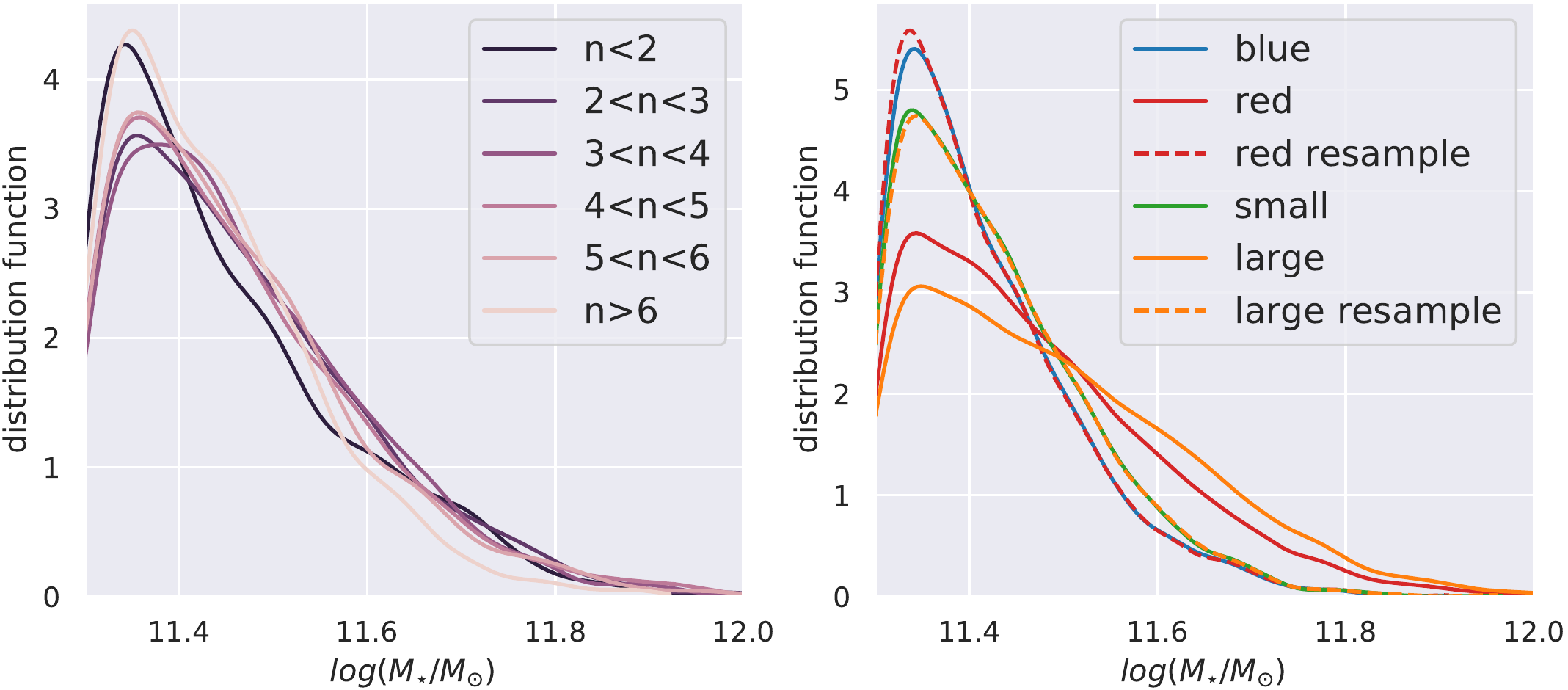}
    \caption{Normalized distribution function
of CMASS galaxies with different morphologies (left), colors and sizes (right) as a function of stellar mass.}
    \label{fig:f1b}
\end{figure*}

As in \citetalias{2021arXiv210911738X}, we use the HSC-SSP PDR2 wide field photometric catalog \citep{2019PASJ...71..114A} as the photometric sample. To obtain more accurate physical properties, we choose sources in the footprints observed with all five bands ($grizy$) to ensure that there are enough bands for SED. Sources around bright objects are masked using {\texttt{\{grizy\}\_mask\_pdr2\_bright\_objectcenter}} flag provided by the HSC collaboration \citep{2018PASJ...70S...7C}. And we use the {\texttt{\{grizy\}\_extendedness\_value}} flag to exclude stars in the sample. Finally, there are around $2\times10^8$ galaxies in our photometric sample. We can also construct a random point catalog ($100/arcmin^2$) with the same selection criteria from the HSC database for PAC analysis. The effective area calculated from the random point number is $501\ deg^2$.

We use the CMASS sample in the Baryon Oscillation Spectroscopic Survey (BOSS; \citet{2012ApJS..203...21A}; \citet{2012AJ....144..144B}) that consists of massive galaxies with $i<19.9\ mag$ as spectroscopic sample (population 1). We first select galaxies in the redshift range of $0.5<z_s<0.7$, and then cross match it with the HSC photometric sample we constructed above and The DESI Legacy Imaging Surveys DR9 catalog \citep{2019AJ....157..168D}. After that, we get the magnitudes in seven bands $grizyW1W2$ for each CMASS galaxy in the footprint of HSC. Then we calculate the physical properties (e.g. stellar mass and rest-frame colors) for these galaxies using the spectral energy distribution (SED) fitting code {\texttt{CIGALE}} \citep{2019A&A...622A.103B}. The details of SED have been discribed in \citetalias{2021arXiv210911738X}. As noticed by previous
studies \citep{2013MNRAS.435.2764M,2016MNRAS.457.4021L,2018ApJ...858...30G}, the CMASS sample is complete
to stellar mass $M_{*}\approx 10^{11.3}{\rm M_\odot}$. Therefore, we adopt a stellar mass cut at $10^{11.3} {\rm M_\odot}$ in this study. Moreover, we only consider central galaxies in the spectroscopic sample, so we select galaxies which do not have more massive neighbors within the projected distance of $1Mpc\ h^{-1}$. Finally, there are $8028$ massive ($>10^{11.3}{\rm M_\odot}$) central galaxies left in the spectroscopic sample.

\begin{figure*}
\plotone{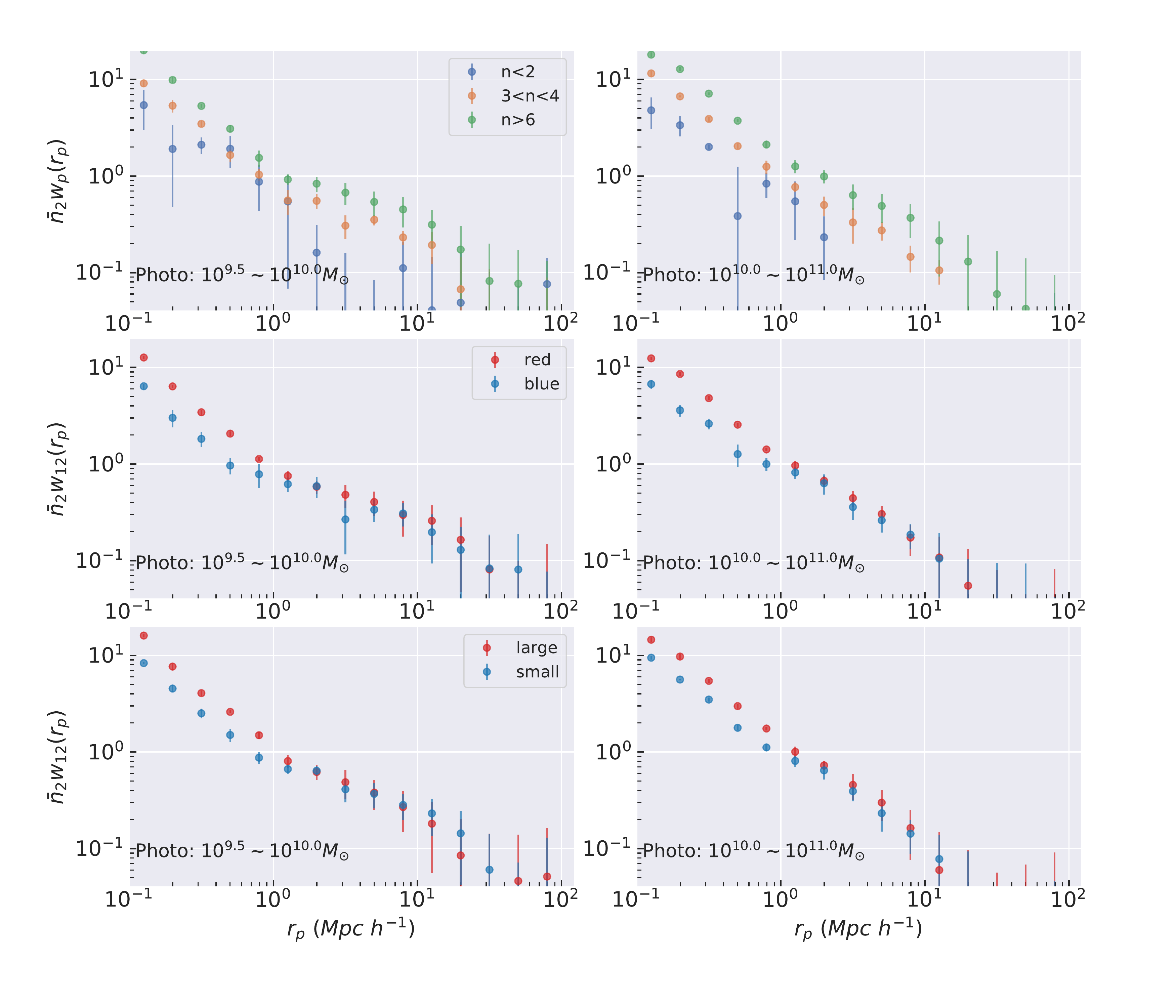}
\caption{Measurements of $\bar{n}_2w_p(r_p)$. Each panel shows the results for the same photometric mass bin and for CMASS subsamples divided according to their morphologies (top), colors (middle) and sizes (bottom). Only three morphology bins are shown for to better illustrate the results.\label{fig:f2}}
\end{figure*}

To study the morphology and size of galaxies, we fit the HSC $z$ band images with the S\'ersic profile \citep{1963BAAA....6...41S} for all the galaxies in our spectroscopic sample. The radial luminosity distribution of a galaxy is described by the S\'ersic form:
\begin{equation}
\label{eq:1}
I(R) = I_e\exp\bigg\{-b_n\bigg[\bigg(\frac{R}{R_e}\bigg)^{1/n}-1\bigg]\bigg\},
\end{equation}
where $b_n$ satisfies:
\begin{equation}
\label{eq:2}
\gamma(2n; b_n)=\frac{1}{2}\Gamma(2n).
\end{equation}
$I_e$ is the intensity at the half-light radius $R_e$, and $n$, called the S\'ersic index, describes the compactness of a luminosity profile. $\Gamma$ and $\gamma$ are respectively the Gamma function and lower incomplete Gamma function. Generally speaking, disk galaxies have exponential profiles with $n = 1$, and ellipticals follow the de Vaucouleurs profiles with $n = 4$. The 2D image fitting is performed using \textsc{galfit} \citep{2002AJ....124..266P} with the point spread function (PSF) taken into account. Other sources falling in the fitting regions are detected and masked out using \textsc{sextractor} \citep{1996A&AS..117..393B}. Since an ongoing merger with companions may lead to a meaningless fitting result with a low $n$, we visually inspect the images of galaxies with $n<2$ and abandon those with apparent major mergers. 

In the left panel of Figure \ref{fig:f1a}, we show the rest-frame $u-r$ color vs. morphology diagram of the CMASS sample. Colors of galaxies is distributed bimodally and weakly correlated with $n$. We adopt $u-r=2.25$ as the color cut for blue and red galaxies with the average S\'ersic index $3.06$ and $4.86$ respectively. The right panel of Figure \ref{fig:f1a} shows the galaxy size $R_e$ vs. morphology diagram, in which the galaxy size slightly increases with increasing $n$. We use the median galaxy size $R_e=9.7\ kpc$ as the dividing line for large and small size galaxies.

In Figure \ref{fig:f1b}, we show the stellar mass distributions for galaxies with different morphologies (left), colors and sizes (right). The mass distribution is nearly the same for galaxies with different morphologies. Therefore, the halo mass difference for galaxies with different morphologies, as we will show in the next section, is independent of the stellar mass. However, large and/or red galaxies tend to be more massive than small and/or blue ones. To eliminate the stellar mass dependence, we construct a controlled sample of large and red galaxies that has the same stellar mass distribution as the small and blue one (dashed lines), and will use it in the following analysis.

\section{Results}\label{sec:result}
Following \citetalias{2021arXiv210911738X}, we calculate $\bar{n}_2w_p(r_p)$ between the photometric sample in 2 mass bins $[10^{9.5},10^{10.0},10^{11.0}]{\rm M_{\odot}}$ and the spectroscopic sample within the stellar mass range $10^{11.3}{\rm M_{\odot}}<M_{\star}<10^{11.7}{\rm M_{\odot}}$ that contains most of the CMASS galaxies. In addition, to study the morphology, color and size dependences, we divide the spectroscopic sample into 6 sub-samples according to $n$, 2 sub-samples according to $u-r$ color or $R_e$. Errors are estimated using jackknife resampling by further dividing each spectroscopic sub-samples into 50 sub-samples. Completeness of the photometric sample is considered using the z-band completeness limit $C_{95}(M_{\star})$ defined in \citetalias{2021arXiv210911738X} (See their Figure 1).

\begin{figure}
\plotone{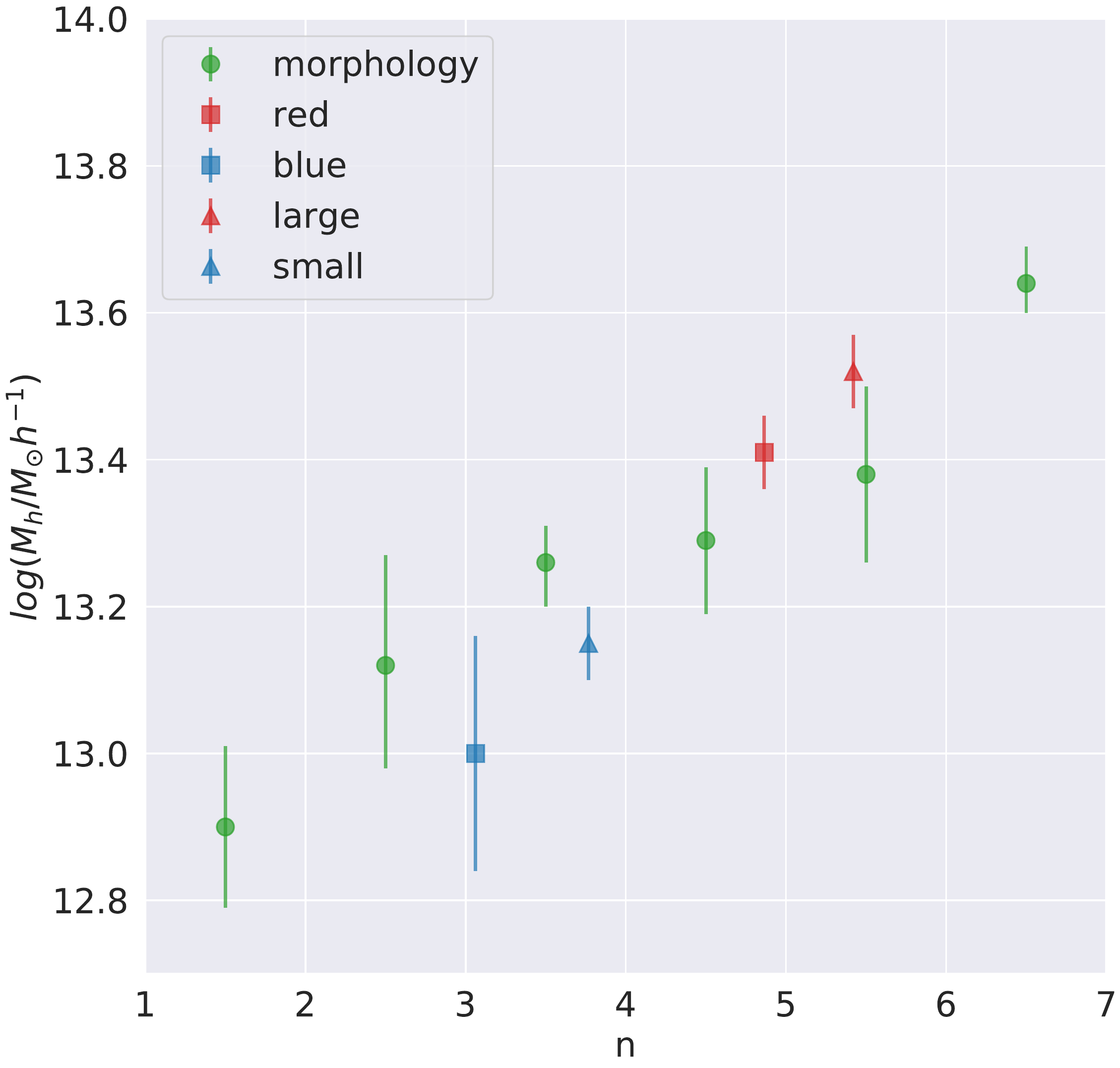}
\caption{Average halo mass estimated from abundance matching for central galaxies with different morphologies, colors and sizes. Green dots show the halo mass for different morphologies, blue and red squares show the halo mass and average $n$ for blue and red galaxies, and blue and red triangles show the halo mass and average $n$ for galaxies with small and large sizes.\label{fig:f3}}
\end{figure}

The measurements are shown in Figure \ref{fig:f2}. Each panel shows the results for the same mass bin but with different morphologies, colors or sizes. To better illustrate the results, we only show the measurements for three morphology bins while the conclusion hold for all six morphology bins. The measurements are overall robust within $0.1\ {\rm Mpc}\ h^{-1}<r_p<10\ {\rm Mpc}\ h^{-1}$. For galaxies with different morphologies, $\bar{n}_2w_p(r_p)$ increases with $n$, indicating that galaxies with more compact structure have more satellites. Since the halo mass of central galaxies is highly correlated with their satellite distribution, it implies that compact galaxies reside in more massive dark matter halos. Similarly, red (large) galaxies have higher $\bar{n}_2w_p(r_p)$ than their blue (small) counterparts, and are expected to have larger host halos.

To quantify the halo mass difference between galaxies with different morphologies, colors and sizes, we use abundance matching (AM, \citet{2010MNRAS.402.1796W,2013MNRAS.428.3121M,2019MNRAS.488.3143B}) to estimate the halo mass of galaxies. As in \citetalias{2021arXiv210911738X}, we use the $\Lambda$CDM CosmicGrowth Simulation \citep{2002ApJ...574..538J,2019SCPMA..6219511J} with cosmological parameters $\Omega_m = 0.268$, $\Omega_{\Lambda} = 0.732$ and $\sigma_8 = 0.831$. The box size is $600\ {\rm Mpc}\ h^{-1}$ with $3072^3$ dark matter particles and softening length $\eta = 0.01\ {\rm Mpc}\ h^{-1}$. The SHMR can be described by a formula of double power-law form:
\begin{equation}
    M_{\star} = \left[\frac{2}{(\frac{M_{acc}}{M_0})^{-\alpha}+(\frac{M_{acc}}{M_0})^{-\beta}}\right]k\,,
\end{equation}
where $M_{acc}$ is defined as the viral mass $M_{vir}$ of the halo at the time when the galaxy was last the central dominant object. We use the fitting formula in \citet{1998ApJ...495...80B} to find $M_{vir}$. The scatter in $\log(M_*)$ at a given $M_{acc}$ is described with a Gaussian function of the width $\sigma$. We populate galaxies to halos using results from \citetalias{2021arXiv210911738X} with $M_0=10^{11.65}{\rm M_{\odot}}h^{-1}$, $\alpha=0.33$, $\beta=2.39$, $k=10^{10.20}{\rm M_{\odot}}$ and $\sigma=0.22$.

With the assumption that the halo mass is completely determined by the satellite distribution \citep{2016MNRAS.457.1208H,2021arXiv210405355W} and with the mock galaxy catalog constructed above, the mean halo mass of central galaxies with specific properties can be estimated through matching the observed excess surface density ($\bar{n}_2w_p(r_p)$) to halos with certain halo mass in simulation. To compare observation with simulation, we define:
\begin{equation}
    \chi^2 = \frac{1}{N_p}\sum_{N_p}\left[\frac{\log(\bar{n}_2w_p(r_p))_{sim}-\log(\bar{n}_2w_p(r_p))_{ob}}{\sigma(\log(\bar{n}_2w_p(r_p))_{ob})}\right]^2\,,
\end{equation}
where $N_p$ is the total number of points over which $\bar{n}_2w_p(r_p)$ is compared. We use the Markov chain Monte Carlo (MCMC) sampler {\texttt{emcee}} \citep{2013PASP..125..306F} to perform a maximum likelihood analysis. 

Halo mass estimated from AM for the spectroscopic sample for different morphologies and colors is shown in Figure \ref{fig:f3}. For morphology (green dots) dependence, halo mass increases monotonically with S\'ersic index $n$, and the most compact galaxies ($n>6$) have the halo mass around $5.5$ times larger than the disk galaxies ($n<2$). For color (squares), red galaxies reside in halos $2.6$ times more massive than those hosting blue galaxies. And for galaxy size. large galaxies reside in halos $2.3$ times more massive than those hosting small galaxies.

\section{conclusion and Discussion}\label{sec:conclusion}
In this paper, we report the morphology, color and size dependences of the SHMR for massive ($10^{11.3}{\rm M_{\odot}}<M_{\star}<10^{11.7}{\rm M_{\odot}}$) central galaxies. Using CMASS and HSC-SSP samples and with PAC developed in \citetalias{2021arXiv210911738X}, we calculate the excess surface density $\bar{n}_2w_p(r_p)$ of satellites and neighbours for the massive central galaxies with different morphologies ($n$), colors ($u-r$) and sizes ($R_e$). We find that, at the same stellar mass, galaxies with more compact morphology, red color and/or large size are surrounded by more satellites. Using AM, we estimate the halo mass for central galaxies with different morphologies, colors and sizes. The results show that more compact, red and/or large central galaxies reside in more massive halos. Specifically, the most compact galaxies ($n>6$) have the halo mass $5.5$ times larger than the disk ones ($n<2$), red galaxies reside in halos $2.6$ times more massive than the blue counterparts, and galaxies with large size have the halo mass $2.3$ times larger than galaxies with small size.

The physical origin of the morphology or color dependence of SHMR is still not well understood. One possible picture is that, for the same halo mass, compact galaxies assemble and transform their morphology earlier and then their star formation is quenched by mechanisms such as active galactic nucleus (AGN) feedback. After the quenching, their halos stills keep growing under the hierarchical formation framework. And for more disky galaxies, the star formation is quenched later or even continues, resulting in a larger stellar mass at a fixed halo mass. Our results also show that the morphology and color dependences are in some way degenerate, and it seems the morphology dependence is more fundamental at least for the massive galaxies. 

The size dependence of SHMR may be explained by the minor merger scenario \citep{2009ApJ...699L.178N,2010MNRAS.401.1099H} that the rate of minor mergers is higher in more massive halos and minor mergers result in size increase. If minor mergers are more frequent in more massive halos, this may introduce a halo mass-size correlation at fixed stellar mass.

In future, with larger photometric and spectroscopic samples, we plan to check whether the morphology, color and size dependences of SHMR hold for lower mass central galaxies. We also plan to verify the estimated halo mass from AM using weak lensing measurements. 

\begin{acknowledgments}
The work is supported by NSFC (12133006, 11890691, 11621303) and by 111 project No. B20019. We gratefully acknowledge the support of the Key Laboratory for Particle Physics, Astrophysics and Cosmology, Ministry of Education. This work made use of the Gravity Supercomputer at the Department of Astronomy, Shanghai Jiao Tong University.
\end{acknowledgments}

\bibliography{sample63}{}
\bibliographystyle{aasjournal}



\end{document}